\documentclass[10pt,letterpaper]{article}
\usepackage[top=0.85in,left=2.75in,footskip=0.75in]{geometry}

\usepackage{amsmath,amssymb}

\usepackage{changepage}

\usepackage[utf8x]{inputenc}

\usepackage{textcomp,marvosym}

\usepackage{cite}

\usepackage{nameref,hyperref}

\usepackage[right]{lineno}

\usepackage{microtype}
\DisableLigatures[f]{encoding = *, family = * }

\usepackage[table]{xcolor}

\usepackage{array}

\newcolumntype{+}{!{\vrule width 2pt}}

\newlength\savedwidth

\raggedright
\usepackage[aboveskip=1pt,labelfont=bf,labelsep=period,justification=raggedright,singlelinecheck=off]{caption}

\makeatletter
\renewcommand{\@biblabel}[1]{\quad#1.}
\makeatother

\usepackage{lastpage,fancyhdr,graphicx}
\usepackage{epstopdf}
\fancyheadoffset[L]{2.25in}
\fancyfootoffset[L]{2.25in}
\begin{document}
\vspace*{0.2in}

\begin{flushleft}
{\Large
\textbf\newline{Early Detection of Influenza outbreaks in the United States} 
}
\newline
\\
Kai Liu\textsuperscript{1,3*},
Ravi Srinivasan\textsuperscript{2},
Lauren Ancel Meyers\textsuperscript{1,4}
\\
\bigskip
\textbf{1} Departments of Integrative Biology and Statistics \& Data Sciences, The University of Texas at Austin, Austin, TX, US\\
\textbf{2} Applied Research Laboratories, The University of Texas at Austin, Austin, TX, US\\
\textbf{3} Institute for Cellular and Molecular Biology, The University of Texas at Austin, Austin, TX, US\\
\textbf{4} Santa Fe Institute, Santa Fe, NM, US\\
\bigskip

* kai.liu@utexas.edu

\end{flushleft}
\section*{Abstract}
Public health surveillance systems often fail to detect emerging infectious diseases, particularly in resource limited settings. By integrating relevant clinical and internet-source data, we can close critical gaps in coverage and accelerate outbreak detection. Here, we present a multivariate algorithm that uses freely available online data to provide early warning of emerging influenza epidemics in the US. We evaluated 240 candidate predictors and found that the most predictive combination does \textit{not} include surveillance or electronic health records data, but instead consists of eight Google search and Wikipedia pageview time series reflecting changing levels of interest in influenza-related topics. In cross validation on 2010-2016 data, this algorithm sounds alarms an average of 16.4 weeks prior to influenza activity reaching the Center for Disease Control and Prevention (CDC) threshold for declaring the start of the season. In an out-of-sample test on data from the rapidly-emerging fall wave of the 2009 H1N1 pandemic, it recognized the threat five weeks in advance of this surveillance threshold. Simpler algorithms, including fixed week-of-the-year triggers, lag the optimized alarms by only a few weeks when detecting seasonal influenza, but fail to provide  early warning in the 2009 pandemic scenario. This demonstrates a robust method for designing next generation outbreak detection algorithms. By combining scan statistics with machine learning, it identifies tractable combinations of data sources (from among thousands of candidates) that can provide early warning of emerging infectious disease threats worldwide.

\section*{Author summary}
Early detection of infectious disease outbreaks enable targeted interventions that prevent transmission and mitigate disease burden. However, we lack rapid surveillance systems for many global threats. This paper introduces a hierarchical statistical method for evaluating diverse data sources and incorporating them into powerful outbreak detection algorithms. We apply the method to design a next generation early warning system for influenza epidemics in the US. By monitoring online Google and Wikipedia search activity for information relating to influenza symptoms and treatment, our algorithm can detect the emergence of seasonal influenza months before the official start of the season.

\linenumbers

\section*{Introduction}
Emerging and re-emerging human viruses threaten global health and security. Early warning is vital to preventing and containing outbreaks. However, viruses often emerge unexpectedly in populations that lack resources to detect and control their spread. The silent Mexican origin of the 2009 pandemic \cite{Mexico2009, Fraser2009Pandemic}, unprecedented 2014-2015 expansion of Ebola out of Guinea \cite{Baize2014Ebola}, and the rapid spread of Zika throughout the Americas in 2016-2017 \cite{Zhang2017zika} highlighted critical shortcomings and the potential for life-saving improvements in global disease surveillance.

Traditionally, public health agencies have relied on slow, sparse and biased data extracted during local outbreak responses or collected via voluntarily reporting by healthcare providers. The 21st century explosion of health-related internet data--for example, disease-related Google searches, Tweets, and Wikipedia term visits--and the proliferation of pathogen molecular data and electronic health records have introduced a diversity of real-time, high-dimensional, and inexpensive data sources that may ultimately be integrated into or even replace traditional surveillance systems. In building 'nextgen' surveillance systems, we face the interdependent challenges of identifying combinations of data sources that can improve early warning and developing powerful statistical methods to fully exploit them. 

Engineers have designed anomaly detection methods for statistical process control (SPC)---including the Shewhart \cite{Shewhart1931}, cumulative sum (CUSUM) \cite{page1954cusum, Lorden1971cusum}, and exponential weighted moving average (EWMA) methods \cite{Roberts1959ewma}---to achieve real-time detection of small but meaningful deviations in manufacturing processes from single or multiple input data streams. When the focal process is \textit{in-control}, these methods assume that the inputs are independent and identically distributed random variables with distributions that can be estimated from historical data. Anomalous events can thus be detected by scanning real-time data for gross deviations from these baseline distributions. 

Biosurveillance systems similarly seek to detect changes in the incidence of an event (e.g., infections) as early and accurately as possible, often based on case \textit{count} data. By adjusting SPC methods to account for autocorrelations, researchers have developed algorithms that can detect the emergence or re-emergence of infectious diseases \cite{Fricker2013}. Such methods have been applied to influenza \cite{Cowling2006flu, Griffin2009flu, Boyle2011influenza, Pervaiz2012flu, Mathes2017}, Ross River disease \cite{Watkins2008ross, Pelecanos2010ross}, hand-foot-and-mouth disease \cite{Li2011handfoot, Zhang2013handfoot, Lai2017handfoot}, respiratory tract infections \cite{wieland2007respiratory, Spanos2012respiratory, Mathes2017}, meningitis \cite{karami2017meningitis}, and tuberculosis outbreaks \cite{Kammerer2013tuberculosis}. These models exploit a variety of public health data sources, including syndromic surveillance, case count and laboratory test data. While they achieve high sensitivity and precision, alarms typically sound once an outbreak has begun to grow exponentially and thus do not provide ample early warning. For annual influenza, CUSUM-derived detection methods applied to Google Flu Trends data sound alarms an average of two weeks prior to the official start of the influenza season \cite{Pervaiz2012flu}.

The Early Aberration Reporting System (EARS) \cite{Hutwagner2003ears} was launched by the CDC in 2000s to provide national, state, and local health departments with several CUSUM-derived methods to facilitate the syndromic surveillance. The BioSense surveillance system \cite{Bradley2005biosense} implements methods derived from EARS to achieve early detection of possible biologic terrorism attacks and other events of public health concern on a national level. Two other surveillance systems, ESSENCE and NYCDOHMH \cite{Bravata2004ESSENCE, Shmueli2010NYCDOHMH}, maintained by United States Department of Defense and the New York City Department of Health and Mental Hygiene, respectively, implement EWMA-based methods for outbreaks monitoring. Most of these systems are univariate (i.e., analyze a single input data source) and consider only public health surveillance data collected during local outbreak responses or via voluntarily reporting by healthcare providers. The time lag between infection and reporting can be days to weeks. Thus, the earliest warning possible for an emerging outbreak may be well after cases begin rising.

Over the last decade, public health agencies and researchers have begun to explore a variety of 'nextgen' disease-related data sources that might improve the spatiotemporal resolution of surveillance. Electronic health records (EHR) systems like athenahealth can provide near real-time access to millions of patient records, nationally, and have been shown to correlate strongly with influenza activity \cite{Santillana2016athena}. Participatory surveillance systems like Flu Near You, which asks volunteers to submit brief weekly health reports, also provide a near real-time view of ILI activity \cite{Chunara2013fny}. However, such data sources may be geographically, demographically or socioeconomically biased, depending on the profiles of participating healthcare facilities or volunteers \cite{Brownstein2017FNYbias}. Internet-source data such as Google Trends \cite{Ginsberg2009gft}, Wikipedia page views \cite{McIver2014wiki, Hickmann2015wiki}, and Twitter feeds \cite{Broniatowski2013twitter} exhibit correlations with disease prevalence, and have been harnessed for seasonal influenza nowcasting and forecasting. However, they have not yet been fully evaluated for early outbreak detection, and may be sensitive to sociological perturbations, including media events and behavioral contagion \cite{Chan2011, Chunara2012}. 

Here, we introduce a hierarchical method for building early and accurate outbreak warning systems that couples a multivariate version of EWMA model with a forward feature selection algorithm (MEWMA-FFS). The method can evaluate thousands of data sources and identify small combinations that maximize the timeliness and sensitivity of alarms while achieving a given level of precision. It can be applied to any infectious disease threat provided sufficient data for the candidate predictors. For novel threats, the candidates may include a wide variety of proxies that are expected to produce dynamics resembling the focal threat (e.g., data on closely related pathogens, other geographic regions, or even social responses to non-disease events).

To demonstrate the approach, we design a multivariate early warning system for seasonal influenza using eight years of historical data (2009-2017) and hundreds of predictors, including traditional surveillance, internet-source, and EHR data. The optimal combination of input data includes six Google and two Wikipedia time series reflecting online searches for information relating to the symptoms, biology and treatment of influenza. By monitoring these data, the algorithm is expected to detect the emergence of seasonal influenza an average of $16.4$ weeks (and standard deviatiation of $3.3$ weeks) in advance of the Center for Disease Control and Prevention (CDC) threshold for the onset of the season. In out-of-sample validation, the model detected the fall wave of the 2009 H1N1 pandemic and the 2016-2017 influenza season five and fourteen weeks prior to this threshold, respectively.

\section*{Materials and methods}
\subsection*{Early detection model}
The MEWMA model is derived from a method described in \cite{Joner2008mewma}. We define one time series as \textit{gold standard}, and one value in the range of the gold standard as the event threshold. Events (outbreaks) correspond to periods when observations in the gold standard cross and remain above the event threshold. We project the timing of events in the gold standard time series onto the candidate time series (predictors). We assume that the data falling outside the event periods follow a multivariate normal distribution $\boldsymbol{F}$ (null distribution) with a mean vector $\boldsymbol{\mu}$ and covariance matrix $\boldsymbol{\Sigma}$ that can be estimated from baseline (non-outbreak) data with equations [\ref{mu}] and [\ref{cov}]:

\begin{equation}
\boldsymbol{\mu} = \mathbb{E}(\boldsymbol{X_T} | y_{\boldsymbol{T}} < \varepsilon)
\label{mu}
\end{equation}

\begin{equation}
\boldsymbol{\Sigma} = \text{Cov}(\boldsymbol{X_T} | y_{\boldsymbol{T}} < \varepsilon)
\label{cov}
\end{equation}
Here, $\varepsilon$ is the value of the threshold defining outbreak events. $\boldsymbol{T}$ are all time points at which observations in gold standard $y$ are below event threshold $\varepsilon$. $\boldsymbol{X}_{\boldsymbol{T}}$ is a matrix of observations from candidate time series at time points $\boldsymbol{T}$.

At each time $t$, MEWMA calculates
\begin{equation}
\textbf{S}_t=
\begin{cases}
\max[\textbf{0}, \lambda(\textbf{X}_t - \boldsymbol{\mu})+(1-\lambda)\textbf{S}_{t-1}], & \text{for}\ t>0 \\
\textbf{0}, & \text{for}\ t=0
\end{cases}
\end{equation}
where $\boldsymbol{X}_t$ is a vector of current observation from candidate time series; $\lambda$ is the smoothing parameter $(0 < \lambda < 1)$; $\boldsymbol{S}_t$ is a weighted average of the current observation standardized around $\boldsymbol{\mu}$ and the previous $\boldsymbol{S}$ statistic. Then the multivariate EWMA test statistic $\boldsymbol{E}_t$ is calculated as
\begin{equation}
\boldsymbol{E}_t = \boldsymbol{S}_t^T \boldsymbol{\Sigma}^{-1}_{\boldsymbol{S}_{\infty}} \boldsymbol{S}_t
\end{equation}
\begin{equation}
\boldsymbol{\Sigma}_{\boldsymbol{S}_{\infty}} = \frac{\lambda}{2 - \lambda} \boldsymbol{\Sigma}
\label{covS}
\end{equation}
The MEWMA signals whenever $\boldsymbol{E}_t$ exceeds a predetermined threshold $h$. That is, the observation at time $t$ deviates significantly from the baseline distribution.

\subsection*{Performance measurement}
Given that our objective is to detect emerging outbreaks early and accurately, we evaluate data based on the timing of alarms relative to the start of events. Only alarms within detection windows are considered as true positive alarms. Specifically, we calculate performance of a candidate system (combination of predictors) as given by
\begin{equation}
P(\boldsymbol{X}, \lambda, h; y) = \frac{1}{N} \sum_{n=1}^{N}(1 - \frac{\Delta T_n} {T_{w}}),
\end{equation}
where $N$ is the total number of events in gold standard, $T_{w}$ is the length of the detection window (e.g., sixteen weeks surrounding the start of an event) and $\Delta T_n$ is the time between the start of the detection window and the first alarm for event $n$. If no alarm sounds during the detection window for event $n$, then $\Delta T_n = T_w$. Performance values range from zero to one. A perfect score of one indicates that alarms consistently sound during the first week of the detection window; 0.5 indicates that alarms occur, on average, right at the start of events; lower values indicate delayed alarms, triggered weeks after the event has begun.  

\subsection*{Parameter optimization}
When implementing MEWMA-FFS, we must estimate the smoothing parameter $\lambda$ and the threshold $h$. The parameter pair $(\lambda, h)$ should maximize the performance of the model while minimizing the number of false positive alarms triggered outside detection windows for actual events.

To constrain the number of false positive alarms, we specify the Average Time between False Signals (ATFS) during the training process. This parameter is the expected number of time steps between signals during non-outbreak periods and is given by

\begin{equation}
ATFS \triangleq \mathbb{E}(t^{**} - t^{*} | \tau_s = \infty),
\end{equation}
where $t^{*}$ denotes the time an initial alarm is triggered; $t^{**}$ is the next time an alarm sounds; $\tau_s$ is the first day of an event, with $\tau_s = \infty$ indicating that an event never occurs. The value of ATFS can be estimated using simulations. We first generate samples from the null distribution (data outside event periods), then use the MEWMA procedure described in \ref{mu} - \ref{covS} to trigger alarms, and finally use the spacing between these false alarms to estimate ATFS \cite{Fricker2013}.

To calculate the optimal parameter pair, we begin with fixing a value of ATFS ($\varphi$). Given a set of time series $\boldsymbol{X}$, this constrains the possible choices for parameter pairs $(\lambda, h)$ to a curve $\Gamma(\varphi, \boldsymbol{X})$. The overarching optimization goal is given by
\begin{equation}
\boldsymbol{X}^*, \lambda^*, h^* = \arg \max_{\{\boldsymbol{X}\subset\boldsymbol{\Omega}:|\boldsymbol{X}|=k, (\lambda, h) \in \Gamma(\varphi; \boldsymbol{X}) \}} P(\boldsymbol{X}, \lambda, h; y)
\end{equation}
where $\boldsymbol{X}^*$ is the optimal combination of time series; $\boldsymbol{\Omega}$ is a set of all candidate time series; $k$ is the pre-determined number of time series in the optimization; $\lambda^*$ and $h^*$ are the optimal parameter pair.

To evaluate parameter pairs $(\lambda, h)$ on the curve $\Gamma(\varphi, \boldsymbol{X})$, we consider values of $\lambda$ between zero and one with a step size 0.1. Since ATFS is monotonically increasing in $h$, this allows us to efficiently find the corresponding approximate value of $h$ using the secant method \cite{Allen1998} with the tolerance value of 0.5 and the maximum number of iterations of 100. We plug each resulting parameter pair into the MEWMA model and measure in-sample performance. The parameter pair maximizing the in-sample performance is chosen for out-of-sample prediction.

\subsection*{Forward feature selection}
To choose the optimal combinations of time series for early warning, we implement stepwise forward feature selection algorithm in combination with MEWMA. We begin with no predictors and test the model performance (in terms of the average timing of early detection) when we add each of the possible candidate predictors on their own. We select the time series that most improves model performance as the first predictor. We then repeat the following until we reach a target number of predictors or the model performance levels off: (1) evaluate each \emph{remaining} candidate predictor in combination with predictors already selected for the system and (2) select the candidate that most improves model performance for inclusion in the system. Formally,

\begin{equation}
\boldsymbol{X}_0 := \emptyset\ \text{and}\ \boldsymbol{X}_{i+1} := \boldsymbol{X}_i \cup \Big\{ \arg \max_{x \in \boldsymbol{\Omega} \backslash  \boldsymbol{X}_i } P(\boldsymbol{X}_i \cup {\{x\}} , \lambda, h; y) \Big\}
\end{equation}
where $X_i$ is a set of selected candidate time series at step $i$; $\boldsymbol{\Omega}$ is a set of all candidate time series; $P(\boldsymbol{X}_i \cup {\{x\}} , \lambda, h; y)$ is the performance metric; $y$ is the gold standard; $\lambda$ is the smoothing parameter, and $h$ is the threshold for test statistic.

\subsection*{Optimizing early detection of influenza outbreaks in the US}
We demonstrate the MEWMA-FFS framework by designing an early detection system for influenza in the US using 2010-2016 data. Using national scale ILINet data as the gold standard (described under \emph{Data} below), outbreak events (influenza outbreaks) are defined as ILINet surpassing a specified threshold for at least three weeks. Candidate predictors are selected to detect the onset of influenza outbreaks as early as possible in a specific number of weeks leading up and following the start of each event.

When selecting candidate predictors, all time series are evaluated using six-fold cross-validation. For each fold, one of the six influenza seasons is held out for testing and the other five are used for training. The candidate model is evaluated by the timing of the alarm relative to the actual start of the event, averaged across the six out-of-sample predictions. To constrain false positives, we set the target ATFS to 20 weeks and then choose optimization parameter pairs $(\lambda, h)$ by running 1000 simulations. To reduce the stochasticity of simulation further, each optimization experiment is repeated 40 times and optimal combination of predictors is determined by the median of their ranks.

After building the early detection systems (i.e., selecting optimal combinations of predictors via MEWMA-FFS), we perform two additional rounds of model evaluation. Since the gold standard and predictor data overlap for only six influenza seasons  (2010-2016), we used this data twice: first, as described above, we use six-fold cross-validation (one season held out) to select optimal combinations of predictors for each model; second, we use three-fold cross-validation (two seasons held out) to compare the performance of different optimized models. We report the timing of alarms relative to the official start of each event, the proportion of events detected (recall), and the percentage of true alarms over all alarms (precision) across the three folds. In preliminary analysis, we found that the length of training data does significantly impact model performance (Fig~\ref{ts_length_performance}). Finally, following model construction and comparison on 2010-2016 data, we further evaluate the performance of the best models in comparison to simpler alternatives using true test data from the 2016-2017 influenza season and the fall wave of the 2009 H1N1 influenza pandemic. 

Since we do not reset $\boldsymbol{S}_t$ to zero following alarms, systems tend to signal repeatedly until the observations return to baseline. Therefore, we track only the timing of the first alarm  during continuous clusters of alarms. MEWMA without resetting saves on computation during FFS optimization, as it allows us to reference a single set of stored null distribution calculations when testing for alarms. That is, if $\boldsymbol{F}$ is the null distribution for all candidate time series, we can compute and save the mean vector $\boldsymbol{\mu}$, covariance matrix $\boldsymbol{\Sigma}$, and $\boldsymbol{S}_t$ statistic with $\boldsymbol{X}_t$, the vector of observations from all candidate time series at time $t$. Given a subset $\boldsymbol{U}$ of candidate time series, the test statistic $E_t$ can be computed by using the pre-computed $\boldsymbol{S}_t$ and $\boldsymbol{\Sigma}$ directly.

\subsection*{Choosing an event threshold and detection window}
To speed up the optimization experiments, we tune the event threshold $\varepsilon$ and length of detection window $T_{w}$. We run optimization experiments using eleven ILINet time series across a range of values for $\varepsilon$ and $T_{w}$ (\nameref{threshold_detectionWindow_performance_comparison}). We constrain the $T_{w}$ so that the start of the window did not precede the lowest observation in the onset of a given outbreak. As in our primary analysis, predictors are selected using 6-fold cross validation and compared via a secondary round of 3-fold cross-validation. We considered ILINet event thresholds ranging from 1\% to 2\% and detection windows ranging from 4 to 20 weeks surrounding the onset of an event and found that a combination of $\varepsilon=1.25\%$and $T_{w}=16$ maximizes the timeliness, precision and recall (\nameref{threshold_detectionWindow_performance_comparison}). 

\subsection*{Assessing the trade-off between run-time and performance}
To evaluate the impact of the ATFS on model performance, we run optimization experiments across ATFS values ranging from 5 to 150. In each experiment, predictors are selected and evaluated through cross-validation as described above. For each ATFS value, we run 40 replicates and record their compute time on the Olympus High Performance Compute Cluster~\cite{olympus}. 

\subsection*{Sensitivity analysis}
To evaluate the impact of the training period duration, we run five optimization experiments following the procedures described above, while varying the length of the training time series from 12 years to 4 years: 2004-2016, 2006-2016, 2008-2016, 2010-2016, 2012-2016. To evaluate the importance of including recent data, we run a series of optimization experiments with variable time gaps between the end of a four-year training period and the beginning of a one-year testing period (\nameref{ts_gap_dataVis}). 

\subsection*{Alternative models}
We compare our optimized early detection algorithms with three simpler alternatives. All three models were fit via 3-fold cross-validation on 2010-2016 ILINet data, with two seasons held out in each round. When computing performance, we follow the methods described above for the MEWMA-FFS model: We consider only the first alarm in each cluster and assume the same objective function, event threshold, detection window, and ATFS.
  
  \emph{Week-based trigger}: The model triggers alarms in the same week of every year. Week 34 maximizes the cross-validated performance.
  
  \emph{Rise-based trigger}: The model triggers alarms as soon as ILINet reports increase for $n$ consecutive weeks. We considered 
  $n$ ranging from 2 to 20 weeks and determined that $n=4$ maximizes the cross-validated performance. 
 \emph{Univariate-ILINet US}: we fit the MEWMA-FFS model using national level ILINet data as the sole predictor.
 
\subsection*{Data}
The method evaluates candidate data sources based on ability to detect events in a designated \emph{gold standard} data source. Throughout this study, we use CDC national-scale ILINet data as gold standard and consider the following five categories of candidate data: (a) ILINet; (b) NREVSS; (c) Google Trends; (d) Wikipedia access log; (e) athenahealth EHR.

  ILINet: The CDC complies information on the weekly number of patient visits to healthcare providers for influenza-like illness through the US Outpatient Influenza-like Illness Surveillance Network (ILINet). Current and historical ILINet data are freely available on FLUVIEW \cite{fluview}. We use weekly percentage of ILI patient visits to healthcare providers on both national and Health and Human Services (HHS) scales (which are weighted by state population). The national scale time series serve as our gold standard data, and both national and HHS data are considered as candidate data sources during optimization from 07/03/2009 through 02/06/2017.

  NREVSS: Approximately 100 public health and over 300 clinical laboratories in the US participate in virologic surveillance for influenza through either US World Health Organization (WHO) Collaborating Laboratories System or the National Respiratory and Enteric Virus Surveillance System (NREVSS). All participating labs issue weekly reports providing the total number of respiratory specimens tested and the percent positive for influenza. These data are publicly available on FLUVIEW \cite{fluview}. Our optimization considers both national and HHS scale time series of weekly percentage of specimens positive for influenza from 07/03/2009 through 02/06/2017.
  
  GT: Google Correlate~\cite{googleCorrelate} and Google Trends~\cite{googleTrends} are freely-available tools developed by Google that enable users to (1) find search terms correlated with user-provided time series and (2) obtain search frequency time series corresponding to user-provided search terms, respectively. We first applied Google Correlate to national scale ILINet data between 01/04/2004 and 5/16/2009 and retrieved the top 100 matches (Table~\nameref{GT_search_terms}). We then applied Google Trends to each of the top 100 search terms to obtain search frequency time series for 07/03/2009 through 02/06/2017. These serve as candidate data sources in our optimization.

  Wikipedia: Wikipedia is widely used as a online reference (nearly 506 million visitors per month) \cite{McIver2014wiki}. Researchers have demonstrated a correlation between US ILINet and time series of access frequencies for English-language Wikipedia articles relating to influenza \cite{McIver2014wiki, Hickmann2015wiki}. Using the Delphi Epidata API \cite{Farrow2016thesis}, we obtained the normalized weekly number of hits for each of 53 influenza-related Wikipedia pages listed in \cite{Hickmann2015wiki} from 07/03/2009 through 02/06/2017 (\nameref{wiki_articles}). 
  
  Athena: athenahealth provides cloud-based services for healthcare providers and manages large volumes of electronic health records data. In collaboration with athenahealth, we obtained the following daily data for approximately 71939 healthcare providers across the US from 07/03/2010 to 02/06/2016: the total number of patient visits, the number of influenza vaccine visits, the number of visits billed with a influenza diagnosis code on the claim, the number of ILI visits, the number of visits ordered a influenza test, the number of visits with a influenza test result, the number of visits with a positive influenza test, and the number of visits with a flu-related prescription. We generated 77 time series total for the following seven variables, each aggregated by week and compiled at the national and HHS scale: (1) ILIVisit---the weekly count of ILI visits; (2) ILI\%---the ratio of the number of ILI visits and the total number of visits; (3) FluVaccine---the weekly count of visits with a influenza vaccine; (4) FluVisit---the weekly count of visits billed with a influenza diagnosis code on the claim; (5) Positive\%---the ratio of the number of visits with a positive influenza test result to the number of visits with a influenza test; (6) FluResult---the number of patient visits with a influenza test result; (7) FluRX---the number of patient visits with a flu-related prescription. 

\section*{Results}

\subsection*{Early detection from single data sources}
\begin{figure}[!ht]
  \centering
  \includegraphics[width=110mm]{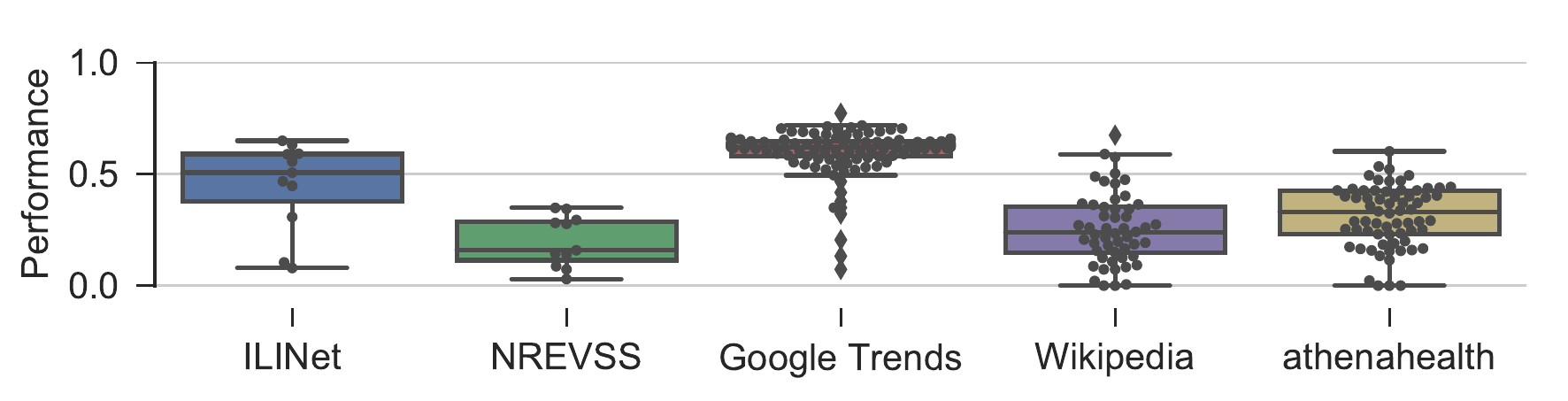}
  \caption{{\bf Early detection by single data sources, summarized by category.}
  For each of the 240 candidate predictors, we fit a univariate detection model and measured performance by averaging early warning across six-fold cross validation (2010-2016). Emergence events for optimization are defined by an ILINet threshold of 1.25\%. The expected performance is highly variable, ranging from 0 to 0.77. A value of one means that the system consistently sounded alarms a full eight weeks prior to the event threshold 1.25\%; a value of 0.5 indicates that, on average, the alarms sound at the time reaching the threshold 1.25\%; lower values indicate delayed alarms.}
  \label{single_performance_threshold125}
\end{figure}

We first fit the early detection model to each of the 240 candidate time series individually and assess their ability to anticipate when ILINet will cross a threshold of 1.25\%. Performance indicates the average timing of alarms based on six out-of-sample tests, with the range of zero to one corresponding to eight weeks after to eight weeks before the event reaching the threshold 1.25\%. The expected performance is highly variable across data sources (Fig~\ref{single_performance_threshold125}), with ILINet and Google source data generally providing earlier warning than laboratory, EHR and Wikipedia data. The Google Trends time series for 'human temperature' provides the best balance of timeliness, precision and recall (Fig~\ref{systems_and_alarms_threshold125}(A)~and~\nameref{alarms_threshold125_supplement}), with an average advanced warning of 14 weeks prior to the CDC's 2\% threshold for the onset of the influenza season \cite{cdcflu}. National scale ILINet data triggers alarms an average of 11.7 weeks prior to the 2\% threshold (Fig~\ref{systems_and_alarms_threshold125}). Several data sources failed to detect any of the seasons, including Wikipedia page views relating to non-seasonal influenza viruses and athenahealth counts of positive influenza tests in HHS regions 8 and 9.

\subsection*{Early detection from multiple data sources}
\begin{figure*}[ht]
  \centering
  \includegraphics[width=\columnwidth]{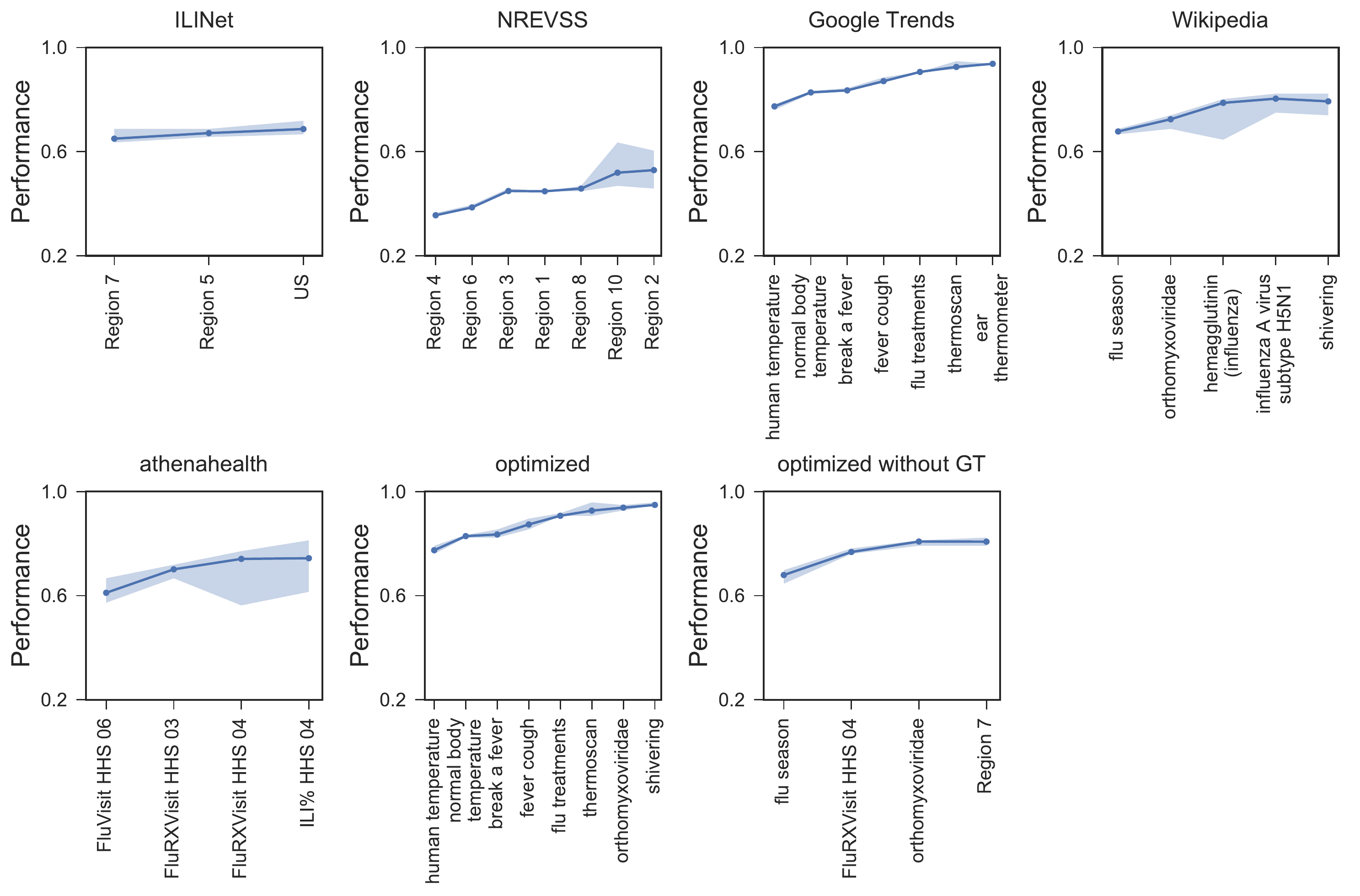}
  \caption{{\bf Performance curves for early detection systems.} Systems were optimized within each data category (ILINet, NREVSS, Google Trends, Wikipedia, and athenahealth) and across all data categories, including and excluding Google Trends. Performance is the average advanced warning within the 16 week detection window surrounding the week when ILINet reaches the event threshold of 1.25\%. Performance equal to one indicates that a model consistently signals eight weeks ahead of the event threshold and zero indicating failure to signal within the detection window. Early detection improves as forward selection sequentially adds the most informative remaining data source until reaching a maximum performance. For the optimal system, the first six predictors are Google Trends sources and the remaining two are Wikipedia sources; for the optimal system excluding Google Trends, the top sources are from Wikiperdia, athenahealth, wikipedia and ILINet, in that order.
  }
  \label{data_selection_curve_threshold125}
\end{figure*}

\begin{figure*}[!ht]
  \centering
  \includegraphics[width=130mm]{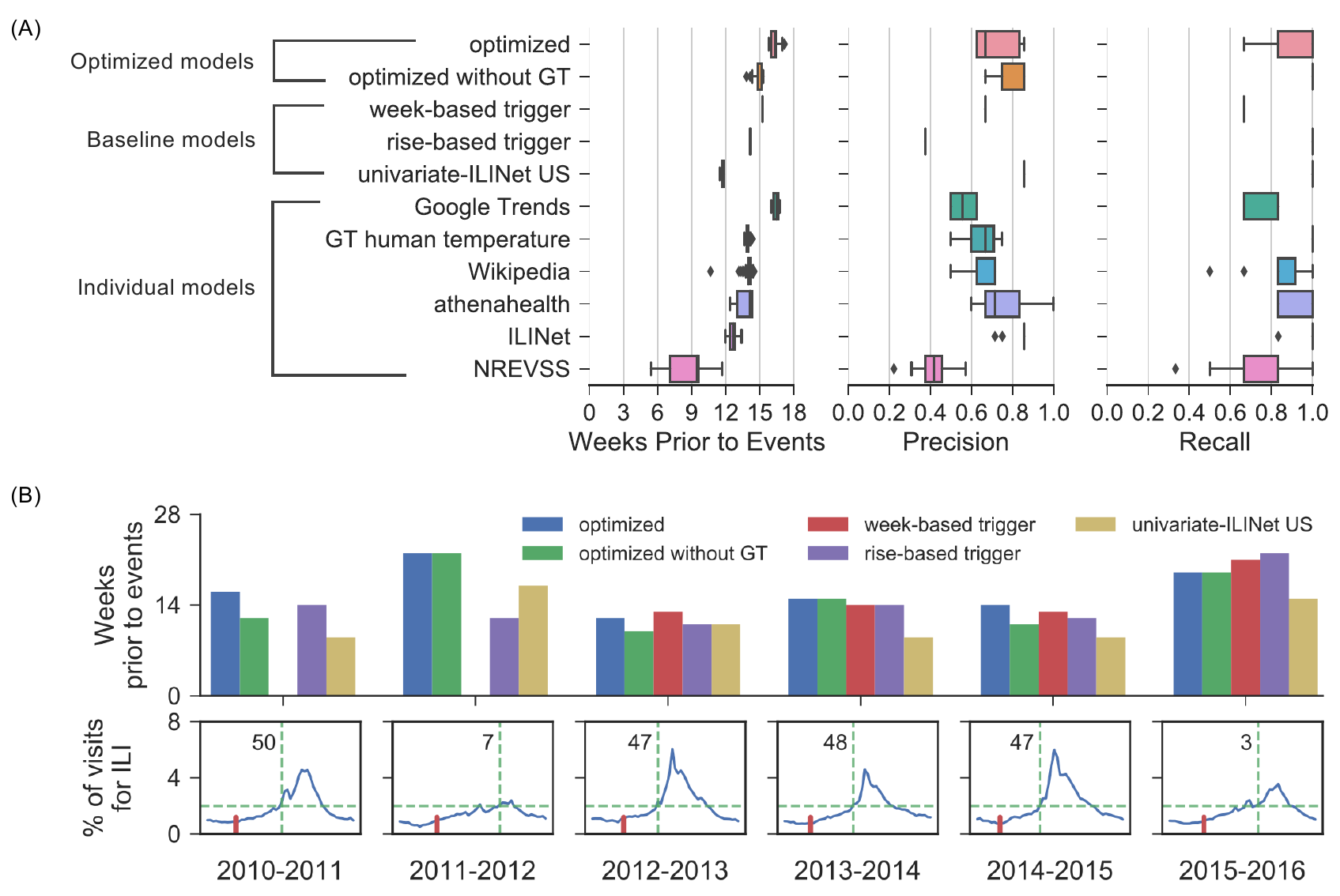}
  \caption{{\bf Performance of optimized US influenza detection algorithms in three-fold cross validation (2010-2016).}
  (A) Distribution of system performance over six influenza outbreaks across 40 replicates, in terms of the timing of true alarms relative to the official onset of influenza seasons (excluding missed seasons), proportion of alarms indicating actual events (precision), and proportion of events detected (recall). (B) Timing of alarms relative to the official onset of each influenza season. Using US ILINet time series (blue curves) as a historical \textit{gold standard}, the detection models were trained to sound alarms as early as possible in the sixteen weeks surrounding the week when ILINet reaches 1.25\%.
  Bar plot (panel 1) shows the advanced warning provided by out-of-sample alarms in terms of weeks in advance of the CDC's 2\% ILINet threshold for declaring the onset the influenza season. Bars not shown indicate missed events. In the lower time series plots, dashed green lines indicate the CDC's seasonal influenza threshold of 2\%; numbers indicate the corresponding week of the year; short red lines indicate the timing of the alarms given by the optimized model.}
  \label{systems_and_alarms_threshold125}
\end{figure*}

We selected optimal combinations of predictors from within each class of data. For CDC ILINet, we considered 11 candidate predictors and found that the optimized system included three time series: ILINet HHS region 7 (Iowa, Kansas, Missouri and Nebraska), ILINet HHS region 5 (Illinois, Indiana, Ohio, Michigan, Minnesota and Wisconsin), and ILINet US (Fig~\ref{data_selection_curve_threshold125}). Across all replicates, HHS region 7 was selected as the most informative predictor, which alone outperforms the optimized system using multiple NREVSS data sources (Fig.~\ref{data_selection_curve_threshold125}). HHS region 9 and US were not selected in all replicates, and just marginally elevate the performance of HHS region 7. Comparing the optimized internet-source systems (Google Trends and Wikipedia) to optimized EHR (athenahealth) system, we find that the best combination of Google Trends time series---\emph{human temperature}, \emph{normal body temperature}, \emph{break a fever}, \emph{fever cough}, \emph{flu treatments}, \emph{thermoscan}, \emph{ear thermometer}---outperforms the others (Fig~\ref{data_selection_curve_threshold125} and \ref{systems_and_alarms_threshold125}(A)).

Across the three-fold out-of-sample tests, the ILINet system detected all six influenza outbreaks with an average advanced warning of 12.7 weeks prior to the CDC's season onset threshold, while the Google Trends system detected 83.3\% of outbreaks (five out of six), with an average advanced warning of 16.4 weeks (excluding missing outbreaks) prior to the official threshold (Fig~\ref{systems_and_alarms_threshold125}(A)~and~\nameref{alarms_threshold125_supplement}). The other systems each detected four to six of six test seasons (not always the same seasons), with average advanced warning ranging from 9.5 to 14.2 weeks (Fig~\ref{systems_and_alarms_threshold125}(A)~and~\nameref{alarms_threshold125_supplement}). Individual ILINet time series generally provide earlier warning than individual EHR and Wikipedia time series. However, performance reverses for optimized multivariate models, with the best ILINet algorithm underperforming both the EHR and Wikipedia algorithms (Fig~\ref{systems_and_alarms_threshold125}(A)~and~\nameref{alarms_threshold125_supplement}).

To build multi-category early detection systems, we applied the optimization method to the 'winners' of the previous experiments. That is, we considered the 26 predictors shown on the first five plots of Fig~\ref{data_selection_curve_threshold125}. The best model includes eight predictors. The top six are all Google Trends: 
\emph{human temperature}, \emph{normal body temperature}, \emph{break a fever}, \emph{fever cough}, \emph{flu treatments}, \emph{thermoscan}; the remaining two are Wikipedia: \emph{orthomyxoviridae} and \emph{shivering}, which only improve the performance of the system marginally (Fig~\ref{data_selection_curve_threshold125}). None of the ILINet, NREVSS, or EHR time series made the cut. The combined system achieves comparable early warning to the optimized Google Trends system while detecting higher proportion of events with lower number of false alarms (Fig~\ref{systems_and_alarms_threshold125}). Furthermore, it sounds alarms earlier than all three alternative models in four out of six seasons. In 2012-2013 all models provide similar early warning; in 2015-2016, the \textit{week-trigger} and \textit{rise-trigger} algorithms signal two and three weeks ahead of our optimized algorithm, respectively (Fig~\ref{systems_and_alarms_threshold125}(B)). The optimized algorithm also produces fewer false alarms than the rise-trigger algorithm and detects a higher proportion of influenza seasons than week-trigger algorithm. (Fig~\ref{systems_and_alarms_threshold125}(B)). The MEWMA model using only ILINet data typically lags all other models in signalling events.

When we exclude Google Trends candidates from optimization, the method selects Wikipedia pageviews of \textit{flu season} as the most informative predictor followed by a combination of EHR, Wikipedia and ILINet time series (Fig~\ref{data_selection_curve_threshold125}). Expected performance declines slightly without Google Trends data. In three-fold out-of-sample evaluation, the six influenza seasons are detected at an average of 14.8 weeks prior to the CDC's 2\% threshold without missing any events (Fig~\ref{systems_and_alarms_threshold125}).

\subsection*{Out-of-sample detection of the 2009 H1N1 pandemic and 2016-2017 influenza season}
\begin{figure*}[!ht]
  \centering
  \includegraphics[width=110mm]{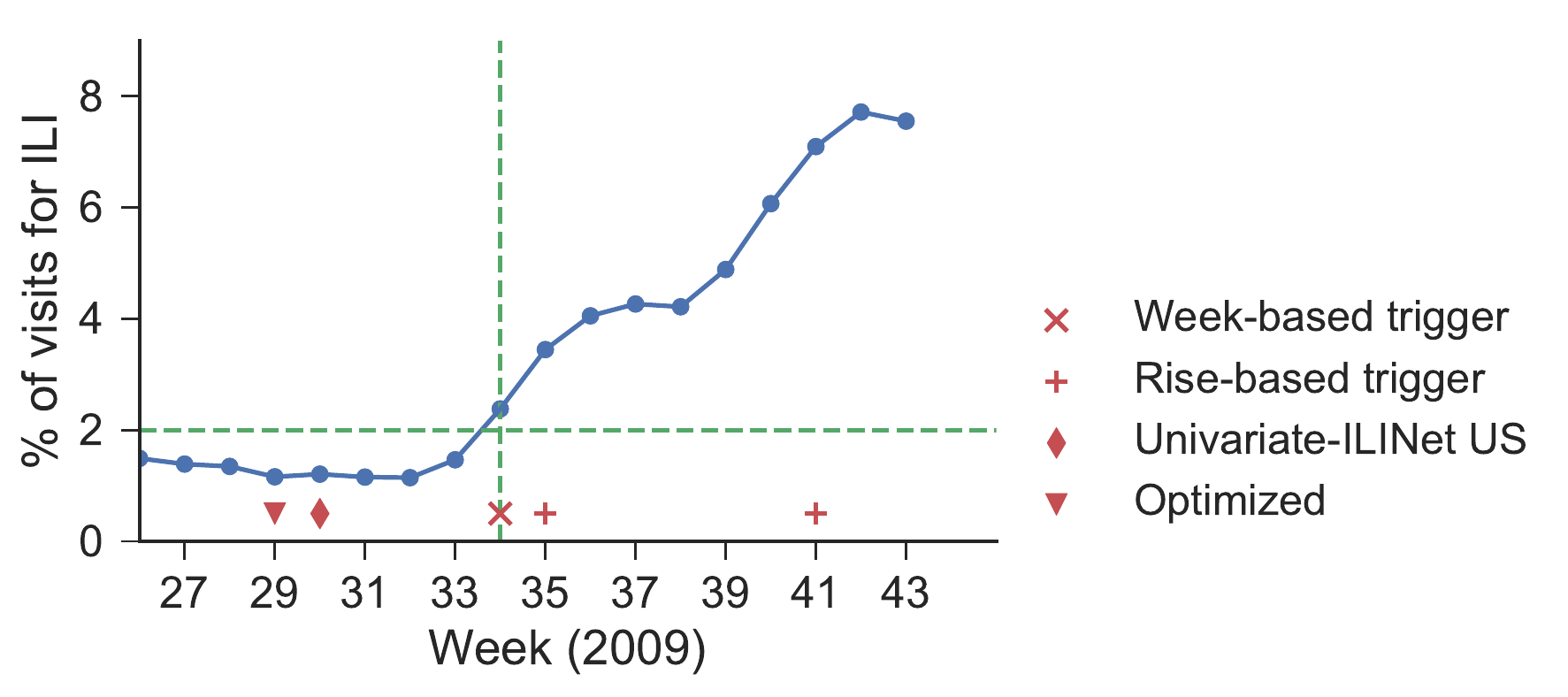}
  \caption{{\bf Early detection of the 2009 H1N1 pandemic (out-of-sample).}
  The optimized model was trained on 2010-2016 ILINet data, and then tested on US ILINet reports (blue curve) during fall wave of the 2009 H1N1 pandemic. It triggered an alarm (triange) five weeks prior to ILINet reaching the official epidemic threshold of 2\% (dashed lines). Red markers indicate timing of alarms triggered by the optimized and baseline models.}
  \label{final_detection_threshold125}
\end{figure*}

We further validated our algorithms using held out ILINet data from two different epidemics. For the 2016-2017 influenza season, the optimized algorithm signaled the start of 2016-2017 season 14 weeks prior to ILINet reaching the CDC's 2\% threshold, which outperforms the univariate ILINet model. However, the week-trigger and rise-trigger algorithms beat the optimized algorithm by two weeks. For the atypical fall wave of transmission during the 2009 H1N1 pandemic, these two models failed to signal the emerging threat. It emerged much earlier in the year than seasonal influenza (thus tripping up the week-trigger algorithm) and at a higher epidemic growth rate (thus outpacing the rise-trigger algorithm) \cite{cdc09pandemic}. The optimal system was able to detect the the fall wave five weeks prior to ILINet reaching the 2\% threshold (Fig~\ref{final_detection_threshold125}). The univariate ILINet model again lags the best model by several weeks in out-of-sample test. This suggests that our optimized multivariate models are more robust for detecting anomalous influenza threats than the simpler alternatives. 

\subsection*{Sensitivity to training period}
\begin{figure*}[!ht]
  \centering
  \includegraphics[width=60mm]{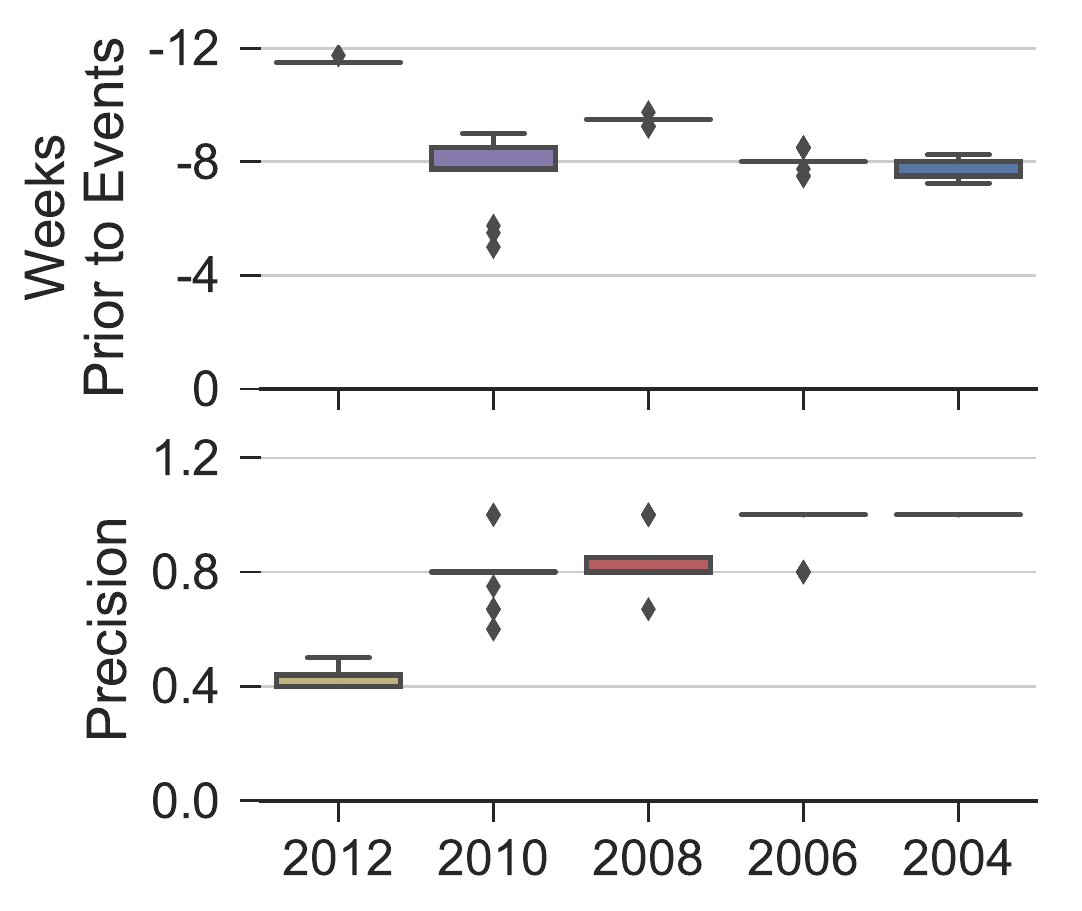}
  \caption{{\bf Duration of training period impacts early detection.} Graphs compare the performance of five systems optimized using continuous training data ranging in length from four to twelve years (each ending in 2016), evaluated via cross-validation on 2012-2016 data. Alarm timeliness (top) unexpectedly declines as the training period increases (maximum likelihood linear regression, P=0.019), while the proportion of true alarms (middle) improves (maximum likelihood linear regression, P=0.000256). Training period does not significantly impact recall (not shown).
}
  \label{ts_length_performance}
\end{figure*}

When we varied the length of the training period from four to twelve years, we selected overlapping sets of optimal predictors, with all five systems including ILINet data for HHS regions 6 and 7 (Table~\nameref{diff_time_window}). The systems detected similar proportions of events. However, the precision (the proportion of true alarms to all alarms) appears to increase with the length of the training period while, surprisingly, the alarms tend to sound later (Fig~\ref{ts_length_performance}). We also found system performance to be fairly insensitive to the gap between the training and testing periods (\nameref{ts_lagged_performance}), suggesting robust performance with only periodic system updates.

\section*{Discussion}
This MEWMA-FFS framework is designed to build robust early outbreak detection systems that harness a variety of traditional and next generation data sources. For seasonal influenza in the US, we identified a combination of freely available internet-source data that robustly detects the start of the season an average of $16.4$ (SD $3.3$) weeks in advance of the national surveillance threshold (ILINet reaching $2\%$). This is five weeks earlier than previously published early detection algorithms based on ILINet and Google data \cite{Cowling2006flu, Pervaiz2012flu}. In a retrospective out-of-sample attempt to detect the fall wave of the 2009 H1N1 influenza pandemic, the optimized multivariate algorithm provided the earliest warning among the competing models. However, it sounded an alarm only five weeks prior to ILINet reaching the national $2\%$ threshold. The shorter lead time may stem from the anomalously rapid growth of the 2009 pandemic. Across the six influenza seasons between 2010 and 2017, ILINet took an average of 9.4 weeks to increase from $1.25\%$ to $2\%$, with a minimum of six weeks in seasons 2012-2013 and 2014-2015; in the fall of 2009, this transpired in a single week (week 34).

Public health surveillance data (e.g., ILINet and NREVSS) can detect emerging influenza seasons on their own, but a combination of eight Google query and Wikipedia pageview time series provided earlier warning across all eight epidemics tested. Although we cannot definitively explain the performance of internet data, we note that 59\% of flu-related Wikipedia English pageviews come from countries outside the US, including the United Kingdom, Canada, and India \cite{McIver2014wiki}. Perhaps earlier influenza seasons elsewhere provide advanced warning of imminent transmission in the US. The utility of Google and Wikipedia data may also stem from their large and diverse user bases and their immediate use following symptoms relative to seeking clinical care\cite{Ginsberg2009gft}. NRVESS is among the mostly costly and time lagged data sources; it performs poorest when considered individually and is never selected for inclusion in combined early detection systems. However, NRVESS provides critical spatiotemporal data for detecting and tracking novel viruses, including pandemic and antiviral resistant influenza, and informing annual vaccine strain selections. Thus, we speculate that NRVESS might rank among the most important sources when designing systems for virus-specific influenza nowcasting and forecasting objectives.

We emphasize that these algorithms are not designed to forecast epidemics, but rather to detect unexpected increases in disease-related activity that may signal an emerging outbreak~\cite{Fricker2013}. Early warning provides public health agencies valuable lead time for investigating and responding to a new threat.
For seasonal and pandemic influenza, such models can expedite targeted public health messaging, surge preparations, school closures, vaccine development, and antiviral campaigns. Influenza forecasting models  potentially provide more information about impending epidemics, including the week of onset, the duration of the season, the overall burden, and the timing and magnitude of the epidemic peak~\cite{shaman2013, brooks2015, ertme2018}. However, they are typically not optimized for early warning or for detecting outbreaks that are anomalous in either the timing or pace of expansion. 

Our conclusions may not be readily applied to influenza detection outside the US or to other infectious diseases. However, the general framework could be similarly deployed to address such challenges. Even for seasonal influenza in the US, our results pertain to only early detection of seasonal influenza activity as estimated from ILINet, and stem from only six seasons of historical data. If we changed the optimization target (i.e., gold standard data) to an EHR or regional ILINet source, the resulting data systems and corresponding performances may differ considerably. Furthermore, as alternative data and longer time series become available, the optimal systems could potentially improve. Early detection systems should therefore be regularly reevaluated and tailored to the specific objectives and geopolitical jurisdictions of public health stakeholders, and our optimization framework can facilitate easy and comprehensive updates. 

This approach requires domain-knowledge in the selection of candidate data sources. Next generation \emph{proxy} data should be relevant to the focal disease and population, such as symptom or drug related search data. Climate and environmental factors may prove predictive for directly transmitted and vector borne diseases, and may be a promising direction for enhancing the early detection systems developed here. This \emph{black box} approach can select data sources with spurious or misleading relationships to the gold standard data. Thus, it may be prudent to screen data sources before and after optimization that are unlikely to correlate reliably with the target of early detection. 
We implemented this MEWMA-FFS framework as an user-friendly app in the Biosurveillance Ecosystem (BSVE) built by the US Defense Threat Reduction Agency (DTRA) \cite{bsve}. Military bioanalysts can now use it to evaluate and integrate diverse data sources into targeted early detection systems for a wide range of infectious diseases worldwide. The versatility of this \textit{plug-and-play} method stems from two assumptions: (1) it simply scans for deviations from underlying distributions rather than modeling a complex epidemiological process, and (2) it does not require seasonality, just historical precedents with which to train the model. We can now more easily harness the growing volumes of health-related data to improve the timeliness and accuracy of outbreak surveillance and thereby improve global health. 


\section*{Supporting information}

\paragraph*{S1 Fig.}
\label{threshold_detectionWindow_performance_comparison}
{\bf Comparison of system performances with different pairs of event threshold $\varepsilon$ and detection window $T_w$ in three-fold cross validation (2010-2016).} Distribution of average system performance over six influenza seasons across 40 replicates, in terms of the timing of true alarms(excluding missed seasons), proportion of alarms indicating actual events (precision), and proportion of events detected (recall).
\begin{figure*}[!hp]
  \centering
  \includegraphics[width=120mm]{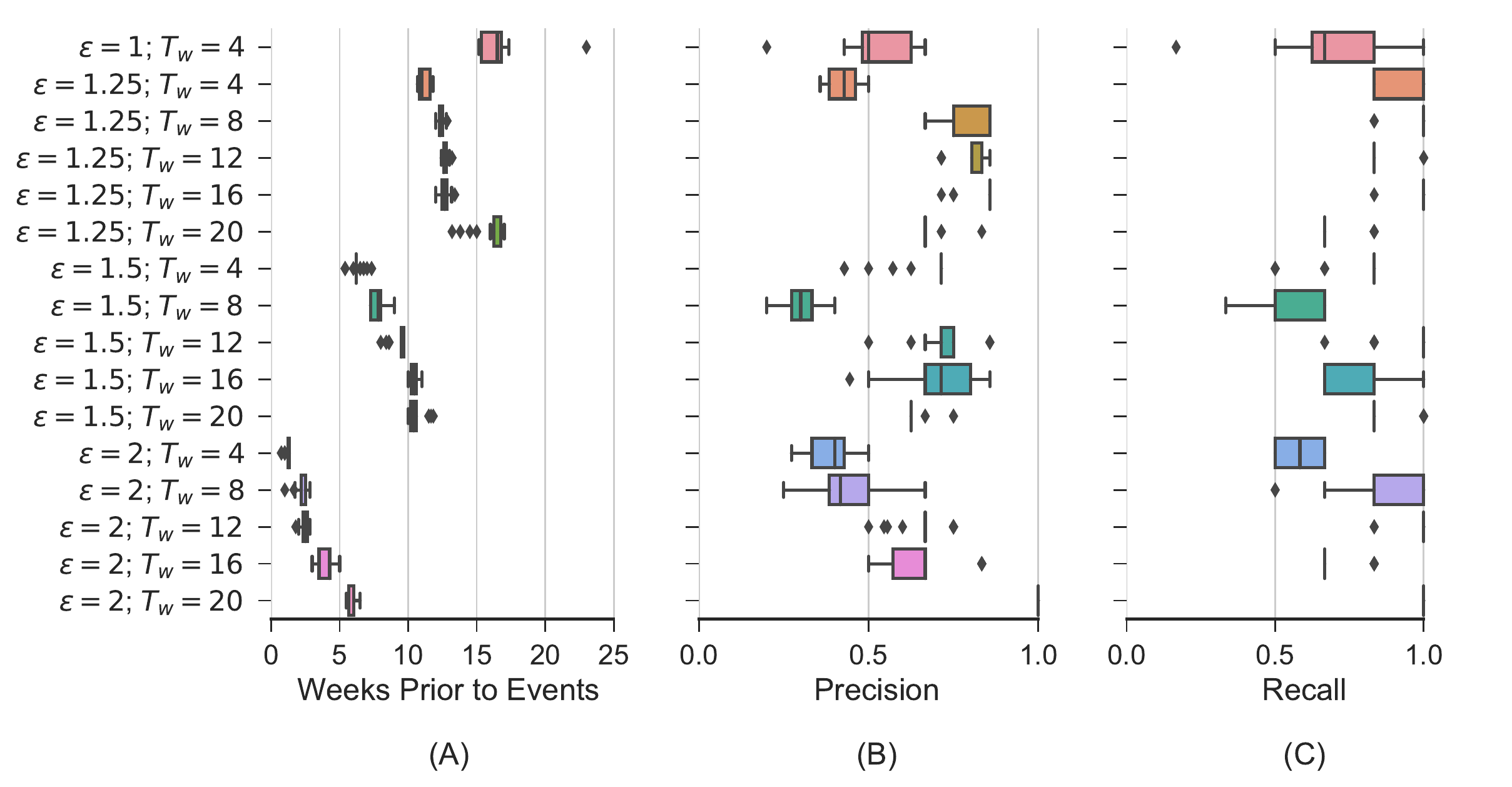}
\end{figure*}

\paragraph*{S2 Fig.}
\label{alarms_threshold125_supplement}
{\bf Out-of-sample detection of US influenza seasons by single source and single category early warning systems.} Using US ILINet time series (blue curves) as a historical gold standard, the detection models were optimized to sound alarms as early as possible in the sixteen weeks surrounding the threshold 1.25\% for optimization. The bar plot (panel 1) shows the alarm timing for each influenza season from 2010-2016 relative to the official ILINet threshold of 2\%. Bars not shown indicate missed events in early detection, while positive values show alarms are triggered prior to the official start of each influenza season. In panel 2, horizontal green dashed lines represent the threshold of 2\%, while vertical green dashed lines indicate the onset of influenza seasons according to the threshold of 2\%; numbers indicate the corresponding week of the year; red short lines show alarm timings for flu seasons from the optimized model.
\begin{figure*}[!ht]
  \centering
  \includegraphics[width=120mm]{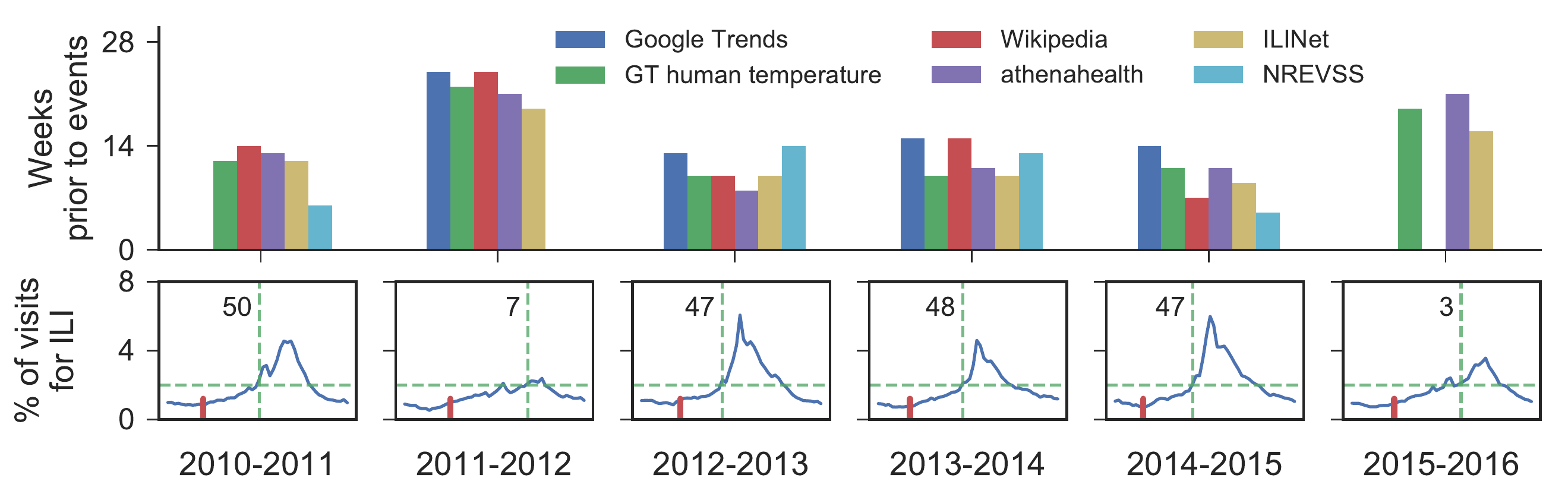}
\end{figure*}

\paragraph*{S3 Fig.}
\label{ATFS_performance_tradeoff}
{\bf The trade-off between timeliness, and precision, recall, running time.} Each system was optimized using different values of ATFS. The three plots show the trade-off between alarms timings and the proportion of alarms indicating actual events (precision), proportion of events detected (recall), and running time of each optimization with 40 repeats running in parallel, respectively.
Each run selected different combinations of predictors (~\nameref{ATFS_tradeoff_table}) and detected influenza emergence an average of 11-14 weeks prior to the official onset of influenza seasons. There is a weak trade-off between timeliness and precision and minimal trade-off between timeliness and recall. The precision is always below 0.9 while recall is equal to one for most of values of ATFS. This is because we consider the timing of only the first alarm in a cluster; the ATFS is expected to impact the total number of alarms but not neccessarily the number of alarm clusters~\cite{Fricker2013}. Meanwhile, a larger value of ATFS requires longer running time for optimization. An optimization experiment with ATFS set to 50 (the value that maximizes timeliness and preceision) requires twice the run time of an experiment using ATFS 20; however, the gain is only one additional week of early warning. Thus, it is valuable to balance performance and compute time when setting ATFS for optimization.

\begin{figure*}[!ht]
  \centering
  \includegraphics[width=\columnwidth]{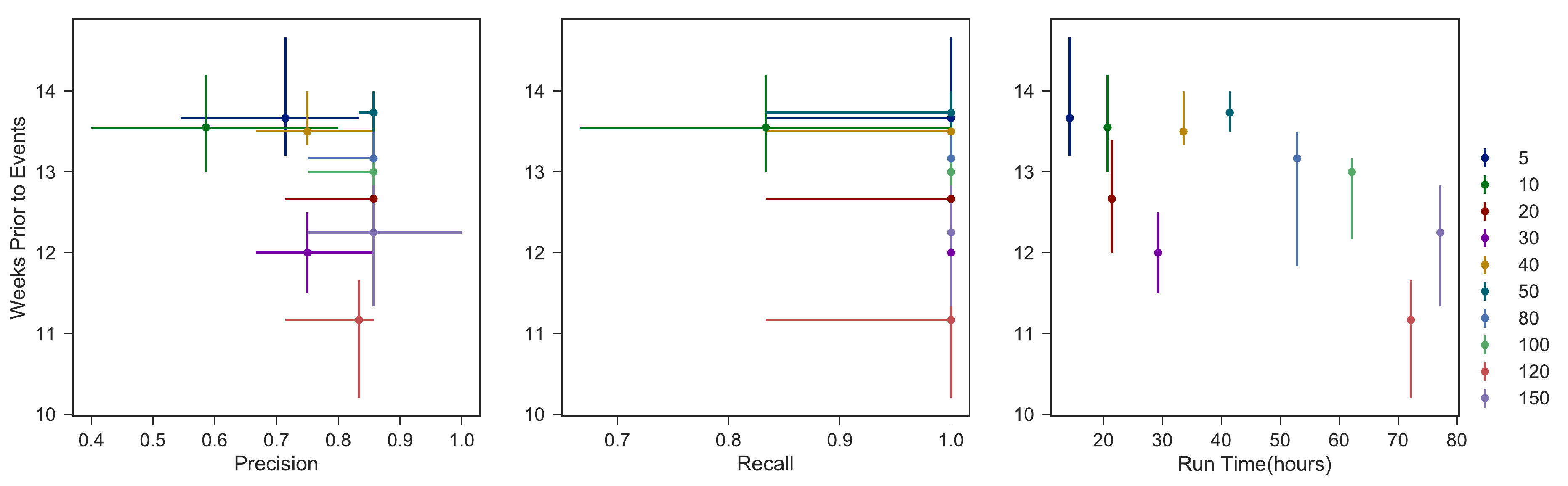}
\end{figure*}

\paragraph*{S4 Fig.}
\label{ts_gap_dataVis}
{\bf Diagram of training and testing periods used in sensitivity analysis.}
\begin{figure*}[!hp]
  \centering
  \includegraphics[width=90mm]{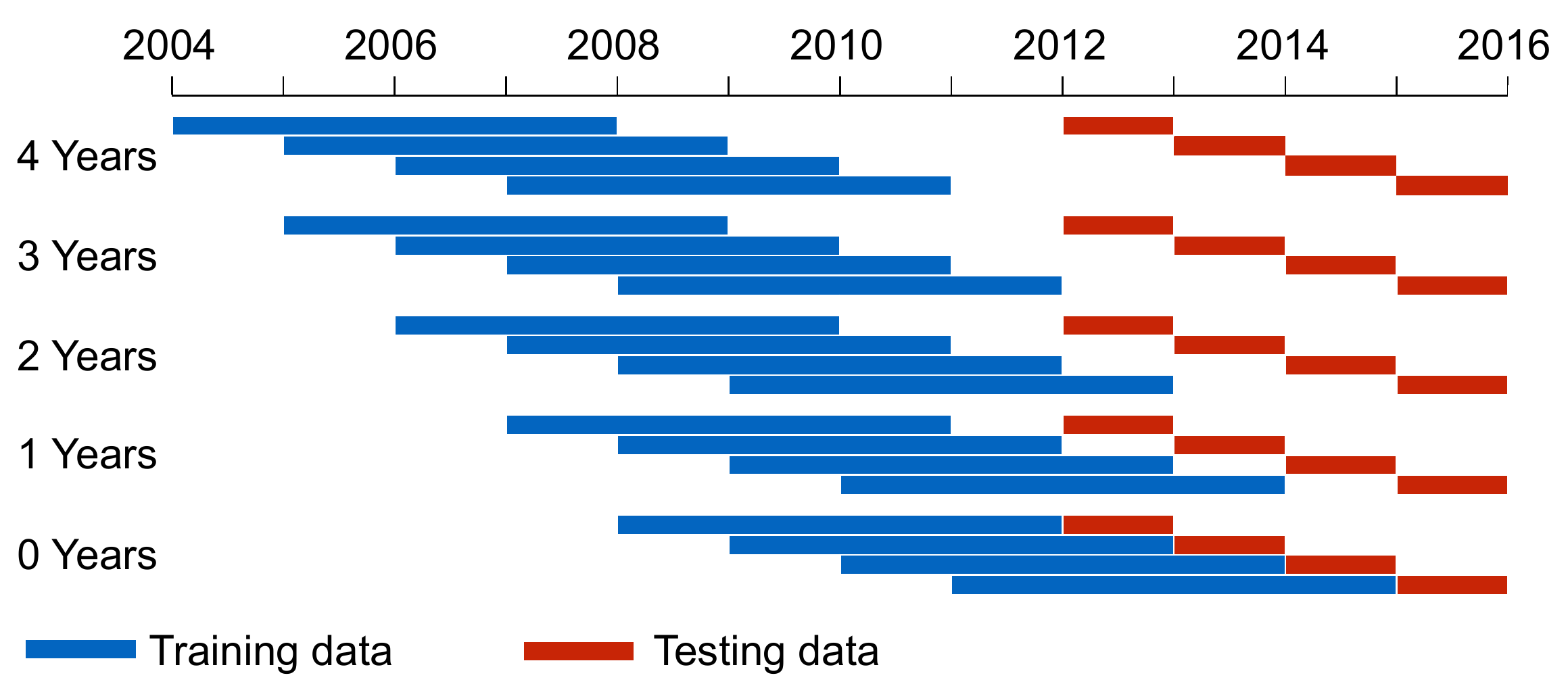}
\end{figure*}

\clearpage

\paragraph*{S5 Fig.}
\label{ts_lagged_performance}
{\bf Sensitivity to the training period.} Each of five systems was optimized using training and testing periods diagrammed in \nameref{ts_gap_dataVis}. The three graphs show performance in terms alarm timing (top), proportion of alarms that correspond to actual events (middle), and proportion of events detected (bottom). Gap between testing and training periods does not appear to significantly impact performance.
\begin{figure*}[!ht]
  \centering
  \includegraphics[width=60mm]{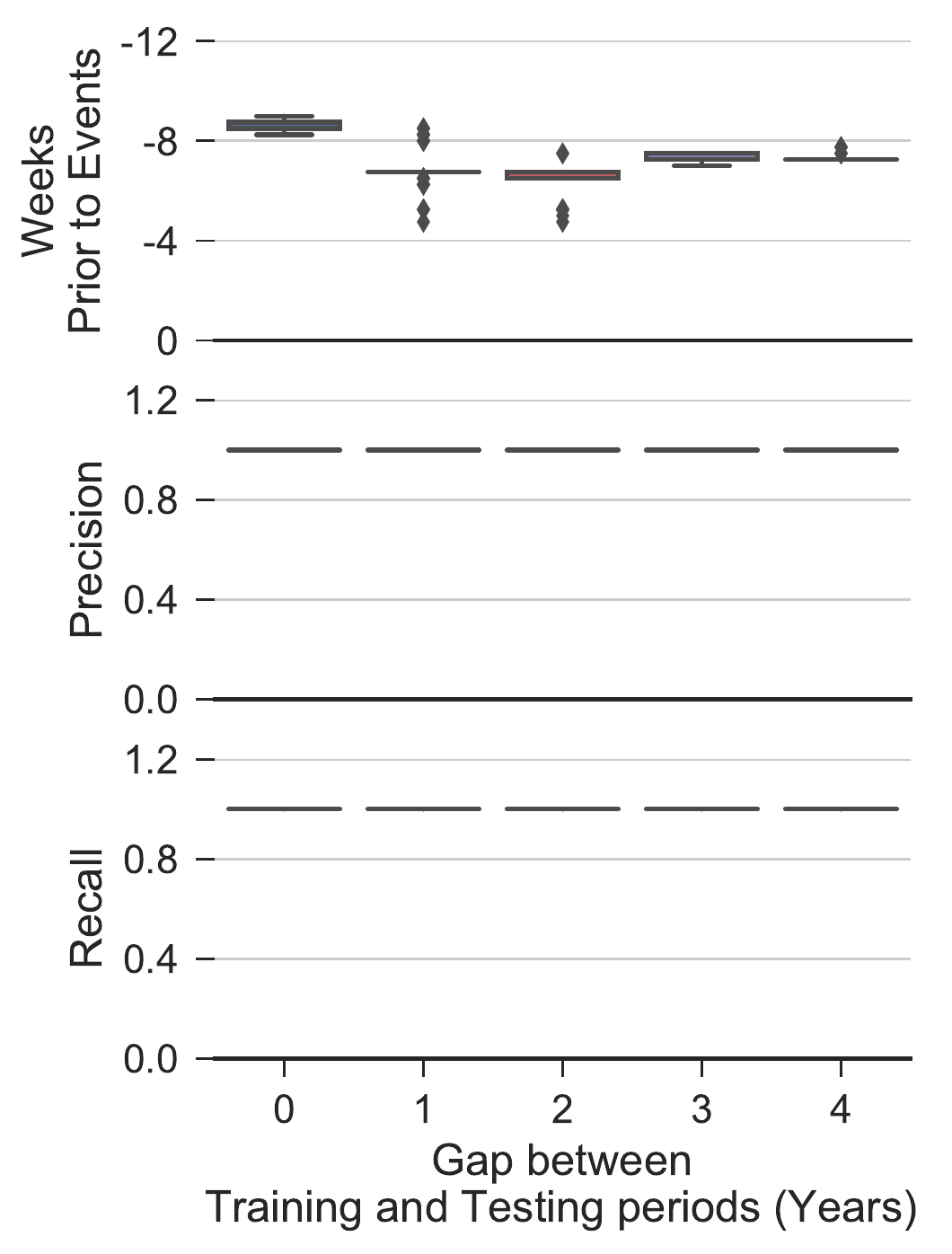}
\end{figure*}

\paragraph*{S1 Table.}
\label{ATFS_tradeoff_table}
{\bf Time series selected for early detection systems across different values of ATFS.} Time series are listed in order of selection, assuming an ILINet threshold of 1.25\% for optimization.
\begin{table}[!hp]
\begin{adjustwidth}{-0.75in}{0in} 
\centering
\begin{tabular}{|l|l|l|l|l|l|l|l|l|l|}
\hline
\multicolumn{10}{|c|}{\bf Value of ATFS}\\ \hline 
5 & 10 & 20 & 30 & 40 & 50 & 80 & 100 & 120 & 150\\ \hline
HHS 7 & HHS 7 & HHS 7 & HHS 7 & HHS 7 & US & HHS 7 & US & US & US\\
 & HHS 6 & HHS 5 & US & US & HHS 4 & HHS 6 & HHS 4 & HHS 9 & HHS 4\\
 & HHS 2 & US & HHS 6 & HHS 10 & HHS 10 & HHS 4 & HHS 6 & HHS 6 & HHS 6\\
 & & & & HHS 4 & HHS 6 & US & HHS 7 & HHS 5 & HHS 7\\
 & & & & HHS 6 & & & & HHS 1 & \\
 & & & & HHS 8 & & & & & \\ \hline
\end{tabular}
\end{adjustwidth}
\end{table}

\clearpage

\paragraph*{S2 Table.}
\label{diff_time_window}
{\bf Data sources selected for early detection systems across variable length training periods.} Time series are listed in order of selection, assuming an ILINet event threshold of 1.25\%
\begin{table}[!hp]
\centering
\begin{tabular}{|l|l|l|l|l|}
\hline
\multicolumn{5}{|c|}{\bf Model Training Period} \\ \hline
2004-2016 & 2006-2016 & 2008-2016 & 2010-2016 & 2012-2016\\ \hline
HHS 5 & US & HHS 7 & HHS 7 & HHS 3\\
HHS 7 & HHS 6 & HHS 1 & HHS 9 & HHS 7\\
HHS 9 & HHS 7 & HHS 6 & HHS 6 & HHS 10\\
HHS 6 & HHS 9 & & & HHS 2\\
HHS 8 & HHS 8 & & & HHS 6\\
 & HHS 2 & & & US\\ \hline
\end{tabular}
\end{table}

\clearpage

\paragraph*{S3 Table.}
\label{GT_search_terms}
{\bf Candidate Google Trends data sources for early detection of seasonal influenza.} Optimization experiments evaluated 100 time series based on each of these search terms.
\begin{table}[!hp]
\begin{adjustwidth}{-1.5in}{0in} 
\centering
\begin{tabular}{|l|l|l|l|}
\hline
\multicolumn{4}{|c|}{\bf Google Search Terms} \\[3pt] \hline
influenza type a & \shortstack[l]{how long is the\\flu contagious} & signs of flu & pneumonia \\[3pt]
exposed to flu & low body & early flu symptoms & flu report \\[3pt]
symptoms of flu & get over the flu & how long does flu last & flu headache \\[3pt]
flu duration & treating flu & normal body temperature & flu cough\\[3pt]
flu contagious & flu vs. cold & get rid of the flu & flu last \\[3pt]
\shortstack[l]{incubation period\\for flu} & flu coughing & break a fever & flu contagious period \\[3pt]
flu fever & having the flu & type a influenza & ear thermometer \\[3pt]
treat the flu & treatment for flu & i have the flu & \shortstack[l]{how to get rid\\of the flu} \\[3pt]
how to treat the flu & human temperature & after the flu & flu how long \\[3pt]
signs of the flu & dangerous fever & when you have the flu & symptoms of bronchitis \\[3pt]
\shortstack[l]{influenza incubation\\period} & cold versus flu & flu in children & \shortstack[l]{what to do if\\you have the flu}\\[3pt]
over the counter flu & the flu & taking temperature & cold and flu \\[3pt]
how long is the flu & remedies for flu & if you have the flu & \shortstack[l]{over the counter\\flu medicine} \\[3pt]
symptoms of the flu & contagious flu & how long flu & flu type \\[3pt]
flu recovery & \shortstack[l]{how long does\\the flu last} & flu germs & treating the flu \\[3pt]
flu and fever & flu lasts & \shortstack[l]{incubation period\\for the flu} & do i have the flu \\[3pt]
flu medicine & have the flu & cold vs. flu & flu care \\[3pt]
flu or cold & oscillococcinum & flu and cold & how long contagious \\[3pt]
is flu contagious & \shortstack[l]{how long is\\flu contagious} & thermoscan & fight the flu \\[3pt]
how long does the flu & flu treatments & flu complications & reduce a fever \\[3pt]
cold symptoms & how to reduce a fever & upper respiratory & fever dangerous \\[3pt]
treat flu & influenza symptoms & high fever & cure the flu \\[3pt]
is the flu contagious & cold vs flu & flu children & medicine for flu \\[3pt]
flu treatment & braun thermoscan & the flu virus & flu length \\[3pt]
flu vs cold & fever cough & how to treat flu &  cure flu \\[3pt]
\hline
\end{tabular}
\end{adjustwidth}
\end{table}

\clearpage

\paragraph*{S4 Table.}
\label{wiki_articles}
{\bf Candidate Wikipedia data sources for early detection of seasonal influenza.} Optimization experiments evaluated 53 time series based on access frequency for each of these Wikipedia articles.
\begin{table}[!hp]
\begin{adjustwidth}{-1in}{0in} 
\centering
\begin{tabular}{|l|l|l|l|}
\hline
\multicolumn{4}{|c|}{\bf Wikipedia Articles} \\[3pt] \hline
Antiviral drugs & Gastroenteritis & \shortstack[l]{Influenza A virus\\subtype H5N1} & Influenza-like illness \\[3pt]
Avian influenza &  Headache & \shortstack[l]{Influenza A virus\\subtype H7N2} & Influenzavirus A \\[3pt]
Canine influenza & \shortstack[l]{Hemagglutinin\\(influenza)} & \shortstack[l]{Influenza A virus\\subtype H7N3} & Influenzavirus C \\[3pt]
Cat flu & Human flu & \shortstack[l]{Influenza A virus\\subtype H7N7} & Malaise \\[3pt]
Common cold & Influenza A virus & \shortstack[l]{Influenza A virus\\subtype H9N2} & Nasal congestion \\[3pt]
Chills & Influenza & \shortstack[l]{Influenza A virus\\subtype H7N9} & Myalgia \\[3pt]
Cough & \shortstack[l]{Influenza A virus\\subtype H1N1} & \shortstack[l]{Influenza A virus\\subtype H10N7} & Nausea \\[3pt]
Equine influenza & \shortstack[l]{Influenza A virus\\subtype H1N2} & Influenza B virus & Neuraminidase inhibitor \\[3pt]
Fatigue (medical) & \shortstack[l]{Influenza A virus\\subtype H2N2} & Influenza pandemic & Orthomyxoviridae \\[3pt]
Fever & \shortstack[l]{Influenza A virus\\subtype H3N8} & Influenza prevention & Oseltamivir\\[3pt]
Flu season & \shortstack[l]{Influenza A virus\\subtype H3N2} & Influenza vaccine & Paracetamol\\[3pt]
Rhinorrhea & Rimantadine & Shivering  & Sore throat \\[3pt]
Swine influenza & Viral neuraminidase & Viral pneumonia & Vomiting \\[3pt]
Zanamivir & & &\\[3pt]
\hline
\end{tabular}
\end{adjustwidth}
\end{table}

\section*{Acknowledgments}
We thank athenahealth, Inc. for providing Electronic Health Records data. Funding was provided by US Department of Defense the Defense Threat Reduction Agency contract HDTRA-14-C-0114, and US National Institute of General Medical Sciences Models of Infectious Disease Agent Study Grant U01GM087719.

\nolinenumbers

%
%
%

\end{document}